\documentclass{article}
\usepackage{arxiv}
\usepackage[utf8]{inputenc} 
\usepackage[T1]{fontenc}    
\usepackage{hyperref}       
\usepackage{url}            
\usepackage{booktabs}       
\usepackage{amsfonts}       
\usepackage{nicefrac}       
\usepackage{microtype}      
\usepackage{lipsum}
\usepackage{graphicx}
\usepackage{xcolor}
\usepackage{amsmath,amssymb}

\title{Instability of social network dynamics with stubborn links}

\author{
  S. Sheykhali \thanks{Corresponding Author} \\
  Department of Physics\\
  University of Zanjan (ZNU)\\
  Zanjan, Iran \\
  \texttt{somaye@ifisc.uib-csic.es} \\
  \And
  A.H. Darooneh \\
  Department of Physics\\
  University of Zanjan (ZNU)\\
  Zanjan, Iran \\
  \And
  G.R. Jafari\thanks{Department of Network and Data Science, Central European University, Budapest, Hungary} \\
  Department of Physics\\
  Shahid Beheshti University \\
  Tehran, Iran\\
}

\begin{document}
\maketitle

\begin{abstract}
This paper studies the signed networks in the presence of stubborn links, based on the structural balance theory. Each agent in the network has a mixture of positive and negative links represent friendly and antagonistic interactions, and his stubbornness about interactions. Structural balance theory affirms that in signed social networks with simultaneous friendly/hostile interactions, there is a general tendency of evolving over time to reduce the tensions. From this perspective, individuals iteratively invert their own sentiments to reduce the felt tensions induced by imbalance. In this paper, we investigate the consequences of the agents' stubbornness on their interactions. We define stubbornness as an extreme antagonistic interaction which is resistant to change. In the current paper we investigated if the presence of stubborn links renders an impact on the balance state of the network and whether or not the degree of balance in a signed network depends on the location of stubborn links. Our results show that a poorly balanced configuration consists of multiple antagonistic groups. Both analytical and simulation results demonstrate that the global level of balance of the network is more influenced by the locations of stubborn links in the resulting network topology than by the fraction of stubborn links. This means that even with a large fraction of stubborn links the network would evolve towards a balanced state. On the other hand, if a small fraction of stubborn links are clustered in five stubborn communities, the network evolves to an unbalanced state.
\end{abstract}


\section{Introduction}
The problem of minimizing social tension or frustration among a group of agents contributes remarkably to the development of interactions, hence shaping communities, alliances, and political groups. Structural balance theory (SBT), proposed by Heider \cite{Heider}, was the first attempt to explain the structure and origin of human tensions in terms of friendship and hostility relationships. This yields that social systems with simultaneous friendly/hostile interactions tend to evolve towards a more balanced situation with minimum stress. Signed networks representing the relation (positive or negative, \textit{i.e.} friendly or hostile) between interacting agents can provide us with significant insights into unveiling the global properties from local interactions \cite{wasserman1994social,easley_kleinberg_2010,Facchetti20953,tan2016evolutionary}. Cartwright and Harary \cite{Cartwright56} formerly modeled SBT to analyze the dynamics and construction of signed networks. Structural balance theory studies the formation, the dynamic features of pairwise interactions among individuals or communities and finally the system's final configuration of interactions.

A generalization of Heider's original hypothesis has been extensively applied to many interesting real-world systems, including social systems \cite{Lerner2016,loreto2,Ware,Kertesz}, the entities within economical activities \cite{cameron2011}, ecologic communities \cite{Saiz2017,ilany2013} and biological neural networks\cite{Zilin}. There has been a large number of previous works studying structural balance in signed networks \cite{aref2017,kirkley2019balance}, using methods motivated by spin glasses \cite{Facchetti,belaza2017}, the dynamic evolution of structural balance \cite{Antal2005,Altafini2012,du2017,Kulakowski1,Hedayati}, reverse transformation of balanced structure \cite{du2018reversing}, time-varying relationship strengths \cite{HASSANI2017} and networks with non-active edges \cite{belaza2019}. 

The SBT states that the network is either balanced or moves towards a balanced state. For the latter case, a measure of balance is needed to trace movement towards balance. Energy function (as specifically defined in Sec. \ref{sec-Model}), stated in terms of cycles is defined to measure the balance degree of the signed network. If a network becomes balanced, it achieves the global minimum of the energy landscape. However, the energy landscape also contains many metastable states, \textit{i.e.}, local minima, named as jammed states \cite{Antal2005}. The concept of such energy landscapes has been studied in a wide range of systems from structural glasses \cite{debenedetti2001} to brain activity \cite{watanabe2014} and social networks \cite{Marvel2009}, to explain how the system gets trapped into local minima as it moves down the energy landscape. Following the SBT, balance dynamics act as the driving mechanism bringing the network to its minimum energy state. In its dynamical evolution, the network may fall into a jammed state in which no single sign change can decrease the energy of the system. Marvel et al. \cite{Marvel2009} observed that the maximum energy allowed for a jammed state is usually found in the lower half of the energy spectrum and it is hard for any system to achieve this upper bound. However, the transformation cost of changing an edge from -1 to 1 is ignored during the dynamical or optimization process. The findings reveal some factors affecting the relationship outside of balance theory in the real world. There is a growing literature studying opinion dynamics in social systems with stubborn agents (whose opinions are kept constant throughout time) \cite{ghaderi2014opinion,yildiz2013binary,Pirani2014SpectralPO,MAstubborn,mirta2014}. Therefore, complete balance is almost never achieved in real systems as a consequence of the complexity of interactions \cite{MAstubborn}. As such, the decisions that people make and the beliefs that people hold have profound consequences on the structure of the network. Here we consider social systems in the presence of "stubborn links", which basically do not follow the network dynamics of attitude change, which is likely to generate tension. We study the stubbornness of negative links, since in accordance with SBT they play a key role in both the structure and dynamics of the network. How does the degree of balance of any signed network depend on the locations of stubborn links? We analyze under which conditions we can avoid certain conflicting situations.

The SBT at most predicts partitioning nodes into one or two strong subsets (bipolar) at the macro-level. This strict condition makes it quite unlikely for a signed network to achieve complete balance in practice. Human social networks are rarely found to be in general strongly balanced (\cite{leskovec2010,van2011micro}). As most empirical structures are not balanced \cite{estrada2014walk}, Doreian and Mrvar have proposed a method for the composition of the network to plus-sets (more than two) as close to balance as possible \cite{doreian1996}. 

In this study, the community formation of stubborn links is observed. Such structures can result in preventing the network from falling into a socially balanced absorbing state. A link-clustering approach to detect communities in terms of closely interrelated links was proposed by Ahn et~al. \cite{Ahn}. Here, we discuss how the so-called stubborn negative links presence results in the system achieving high energy of a jammed state. By means of the energy function and the signed network topology, we address the relationship between them in the presence of such stubborn links. Moreover, we study how many negative subgraphs exist on a network and how does this number depend on the number of nodes. Through simulations, we show that a network can non-trivially tolerate a large population of stubborn links depending on their distribution. Still, the network with a large number of stubborn links does not have a high probability to reach a balanced state. Finally we study how these links affect the energy of the network in terms of imbalanced triads clusters and distribution of community sizes.

\section{Structural Balance with stubborn links}
\label{sec-Model}
\begin{figure}[h!]
	\begin{center}
		\includegraphics[width=0.45\textwidth]{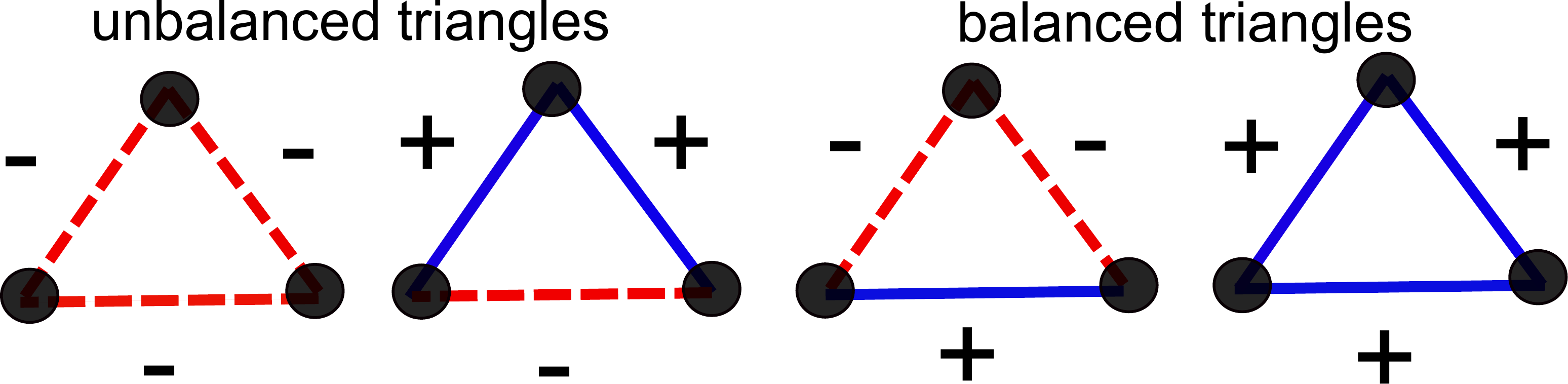}
		\caption{Different balanced and unbalanced configurations of a
			triangle. Solid lines with '+' label represent positive edges and dashed lines with '–' label represent negative edges.}
		\label{figsbt}
	\end{center}
\end{figure}

Following structural balance theory, the two signed triadic configurations (+ - -) and (+ + +) are considered as balanced and (+ + -) and (- - -) signed triadic configurations are considered as unbalanced. The two unbalanced triangles are identified as a source of inherent tension which may drive them into more balanced configurations (Figure. \ref{figsbt}).
We begin by modeling a fully connected network as a signed complete graph $G(N; S)$, with $N$ agents and $S$ edges. Each positive or negative edge $S_{ij}$ representing a feeling of trust/mistrust, like/dislike, or friendship/enmity between the node $i$ and $j$ \cite{Szell,Facchetti}. Following the SBT, a potential energy function to quantify the degree of balance of any signed social network is defined as\cite{Marvel2009}:
\begin{eqnarray}
E=\frac{-1}{\binom{N}{3}}\sum S_{ij}S_{jk}S_{ki}.
\label{e1}
\end{eqnarray}
where $S_{ij}$, $S_{jk}$, and $S_{ki}$ are signed links forming a triad and the sum is overall triangles ${i, j, k}$ of the network. The energy ranges between $-1$ and $+1$ corresponding to a balanced and an imbalanced structure respectively.

Structural balance theory provides the basic components of a dynamic model by specifying sign changes of links in the unbalanced triples to be balanced—a graph is balanced if all of its triples are balanced. Two internally friendly cliques with mutual antagonism between them, is the only sign configuration in global minima (The paradise configuration is considered as the extreme in which one clique is empty). 

\subsection{Stubborn links}
There is no guarantee, however,  that all of the links in the network will cooperate in balance dynamics. Some links may refuse to switch and lead the network to an imbalanced final state (stubborn links). In this term, a network may fall into a final state containing imbalanced triads (Figure. \ref{figsbt}). Changing the sign of any such negative links would thereby render all the triads imbalanced. To create the network, we first generated a matrix of $N$ nodes. The signs "+" or "-" are randomly assigned to the links. The negative stubborn links ($n_{s}=1: N(N-1)/2$) are randomly distributed. Generally, we observe that the presence of stubborn links affects the final balance state, and the deviation depends not only on the fraction but specifically on the location of the stubborn links in the network structure.

\subsection{Stubborn Communities}

\begin{figure}[h!]
	\begin{center}
		\includegraphics[width=0.5\textwidth]{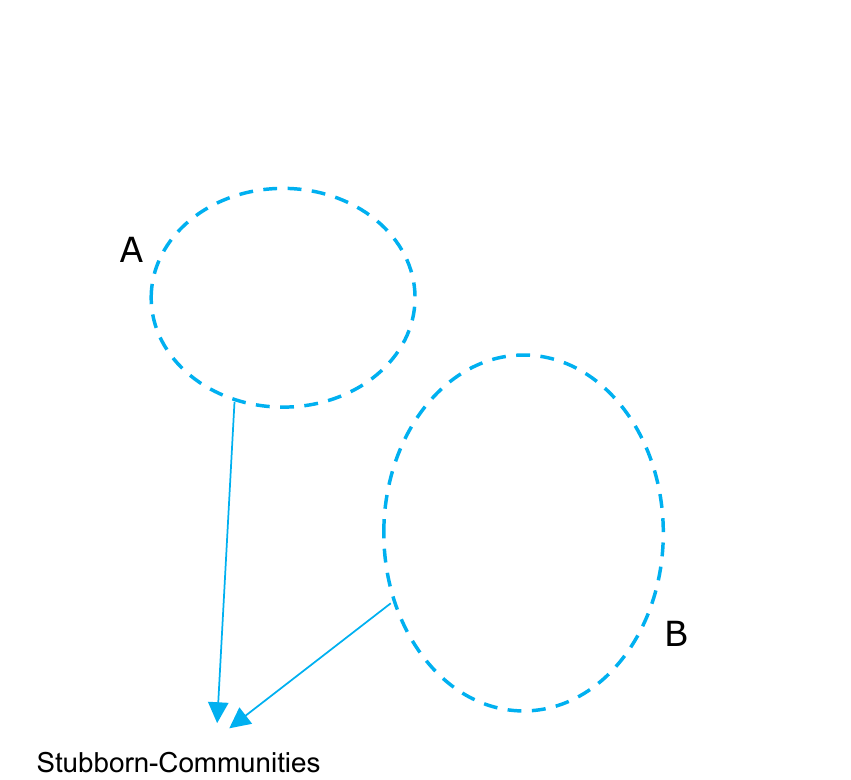}
		\caption{Schematic pattern of a network with stubborn links (red dash lines) clustered in stubborn communities (blue dashed circles); they provoke a stubborn-community, i.e. stubborn links  forming a cluster.}
		\label{St-Com}
	\end{center}
\end{figure}

\begin{figure}[h!]
	\centering
	\includegraphics[width=0.9\linewidth]{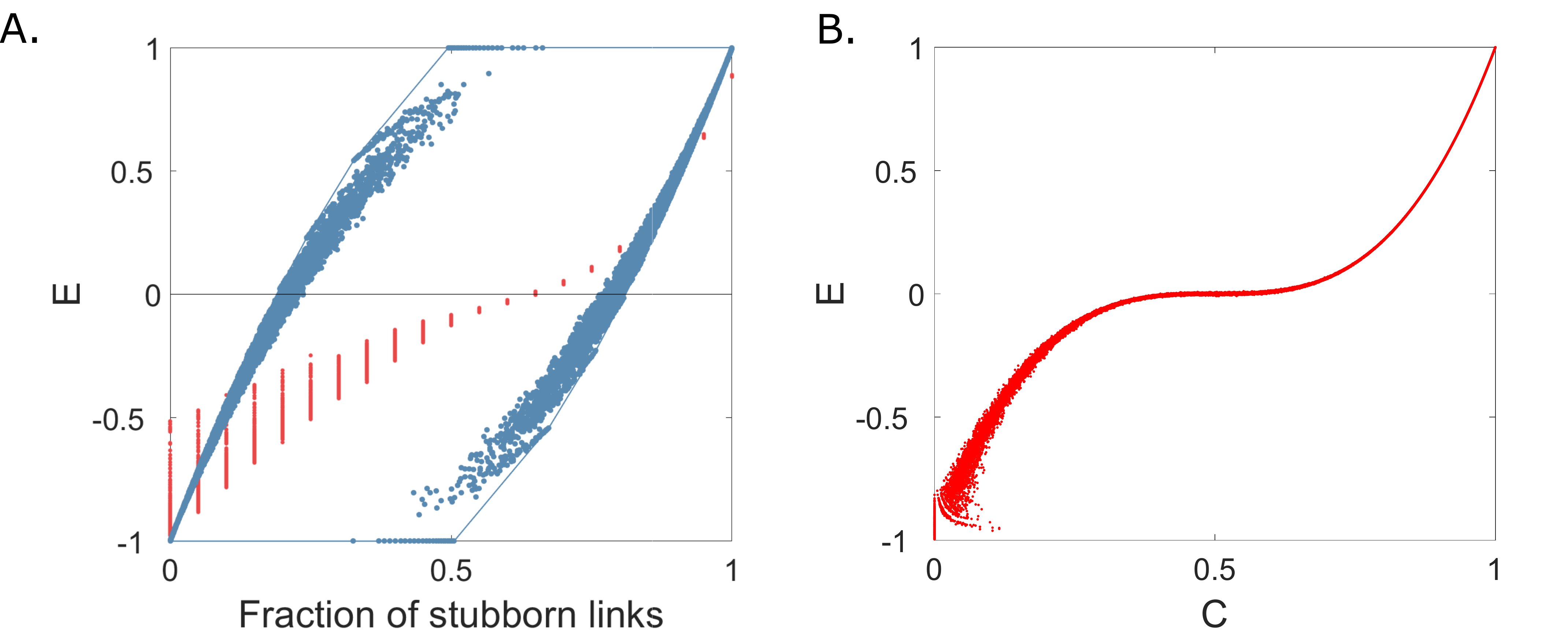}
	\caption{\textbf{(A)} Energy as a function of the fraction of stubborn links for a network with $N=100$ nodes. Each point represents the final state of a single realization. The red points represent the energy reached by networks with a random displacement of stubborn links in the network. The blue points represent the extreme values of the energy of the final state when stubborn links form communities. The blue solid lines are related to a network with approximately same size stubborn communities, the upper line represents the communities formed by the minimum fraction of stubborn links and the lower line correspond to the maximum fraction of stubborn links. \textbf{(B)}  Energy as a function of the clustering coefficient (C) for stubborn links in networks in the final state of every simulation.}
	\label{fig2}
\end{figure}

In a balanced system, the individuals can be partitioned into two or more subgroups \cite{Davis1967} or balanced cliques to minimize the structural conflicts. It has been shown natural modular structure for networks in jammed states \cite{Marvel2009}. The states with high-energy local minima must be structurally more complex than low-energy ones. The communities formed by nodes with only negative or very few positive links have been studied before \cite{Shen}. We define stubborn communities as those cliques made of stubborn links only and having positive links with the rest of the network. Our study shows that when stubborn communities are formed (Figure. \ref{St-Com}), a few number of stubborn links can substantially make the energy positive.
It can be safely said that a tendency to form community structure (community by stubborn links) is present. This can be quantified with the clustering coefficient, $C$ Which is defined as the probability that two incident edges are completed by a third edge to form a triangle number of closed triangles \cite{newman2002random}:
\begin{equation}
C = \frac{ \{ number \; of \;closed \; traingles \} }{ \{ number \; of \; all \;open \; and \; closed \; triplets \}}  
\end{equation}
In order to measure the density of stubborn communities, we use the clustering coefficient measure for negative sign link configurations. Figure.~\ref{fig2}.B shows the scatter plot of energy against the clustering coefficient. What we observe in this Figure is that the formation of stubborn communities moderately grows with energy, while it remains fairly constant in $E=0$ for $0.3<C<0.7$ and again grows to  $C=1$ as unbalanced communities grow in numbers.

Here, the optimization problem is to minimize the fraction of stubborn links necessary to reach an imbalanced state of a network. Starting from an initial random sign configuration, we select and switch single link signs until the network falls into a local minimum. In order to have a statistic of the final states, in every 1000 independent realizations, we embed stubborn links randomly in a  fully connected sign networks, with fraction varying from 0 to 1, and calculated the energy. The red points in Figure. \ref{fig2}.A, show the energy of the final state versus the fraction of stubborn links when they are randomly displaced. Stubborn links clustering is crucial for the definition of the final state network energy, as shown in Figure.~\ref{fig2}.B. Although, a high fraction of such links does not imply a high clustering coefficient. There is a minimum fraction of such links able to move the network to a positive energy final state. In this case, we define the energy upper bound as the maximum energy reachable by the system with this configuration.
On the other side there is also a maximum number of stubborn links able to keep the system into negative energy final states. We define the range of variation of the energy function with respect to these extremes and the energy lower bound as the minimum energy reachable by this configuration.

Without stubborn links, the network eventually reaches a balanced state. By imposing a subset of stubborn links the energy changes significantly. A small fraction of stubborn links has more chance to locate randomly in the network in such a way that they do not link to any other stubborn link. With few stubborn links (less than $20\%$ of all the links) the chance of reaching a balanced state is still high (Figure. \ref{fig2}). However, by increasing the number of stubborn links many imbalanced triads cannot reach a balance, resulting in a frozen state. In this case, the only way to reduce the energy of the system would be by flipping a stubborn link, which does not occur by definition. We aim to determine the optimal placement of the minimum number of stubborn links with the maximum impact on the energy.

The spatial distance apart stubborn links allowing more triangles to be unbalanced. This feature cause more (- + +) unbalance triangles and also increases energy. Progressively increasing the fraction of stubborn links, they less and less likely locate separately and the probability of connecting each other increases. Which the number of (- - +) balance triangles increases and also energy can decreases. On the other hand, more stubborn links, more probability of clustering, more probability of reaching high energy (red points in \ref{fig2}.A). When stubborn links assemble in negative communities, the number of (- + +) unbalance triangles increases causing a raise of energy (upper blue points in \ref{fig2}.A). The lower blue points in \ref{fig2}.A show the extremely opposite state with positive links inside communities and stubborn negative links between them.  We quantify these observations by measuring the clustering coefficient ($C$) of the negative stubborn links in the network, which shows how they progressively assemble into communities (\ref{fig2}.B).

\begin{figure}[h!]
	\begin{center}
		\includegraphics[width=0.9\linewidth]{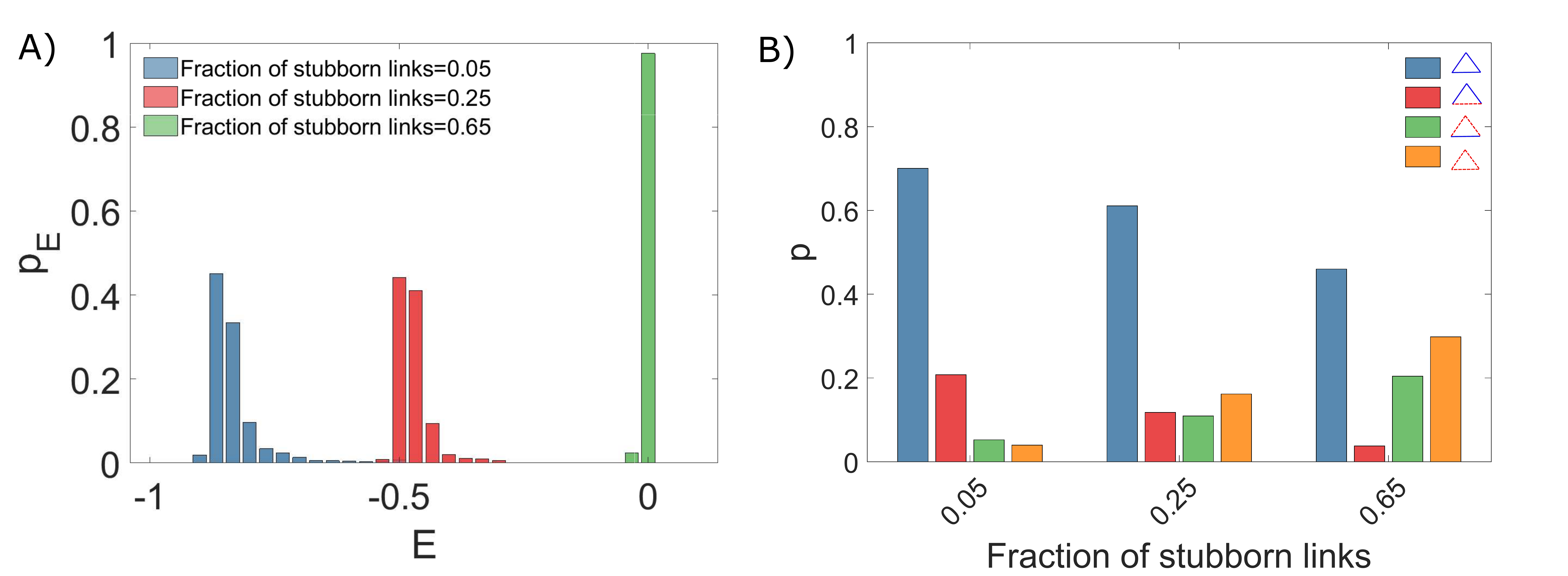}
	\end{center}
	\caption{\textbf{(A)} Distribution of energy of the final state. \textbf{(B)} Distribution of number of triangles for any of the four types. For networks with $N=100$ nodes and $1000$ realizations.}\label{fig3}
\end{figure}

For energy reduction, the distribution of balanced and unbalanced triangles should be taken explicitly into account. In order to visually compare how the energy and the number of balanced and unbalanced triangles change as the fraction of stubborn links increases, we plot in Figure.~\ref{fig3} the distribution of triangles and the distribution of energy. As expected, energy increases and the range of energy decreases due to stubborn links higher density. This is related to the fact that increasing the fraction of stubborn links causes more unbalanced (- - -) triangles to appear.

The tendency of a network over time to become more balanced leads to a reduction in the number of imbalanced triangles, which in this case is followed by a reduction in the number of negative links. On the other hand, to find the minimum number of negative links to allow a positive energy final state we should look at the configurations with more positive links. It ultimately could turn the system into an unstable state in which there is no propensity to reduce the unbalanced triangles in favor of balance \cite{glassy}. Our results show that stubborn links forming communities may lead the system to fall into a local minimum with positive energy. Even a small fraction of stubborn links may evolve into stubborn communities. 

\section{OPTIMUM NUMBER OF Stubborn-Communities}

Marvel \textit{et al.} \cite{Marvel2009} showed that jammed states may be derived from the undirected Paley graphs \cite{bollobas2001}. Our first result is that local minima of clustered networks with stubborn links can have energies above zero (defined in Eq.~\ref{e1}). To see this, note that in a balanced state, there are at least as many unbalanced triangles as balanced triangles. A triad is balanced if any three nodes from any three distinct stubborn-communities are selected (+ + +) or two nodes from one community and the third one from another community (+ - -). On the other hand, three nodes from the same community form an unbalanced triangle. There are therefore $ \sum_{i} \binom {m_{i}}{2} (N-m_{i})+ \binom {m}{3}$ balanced triangles and $ (N-m_{i}) \binom {m_{i}}{3}$ unbalanced triangles, where $i=1,...,k$ is the number of cliques and $m_{i}$ is the number of vertices in the clique i.

\begin{figure}[!h]
	\begin{center}
		\includegraphics[width=0.4\textwidth]{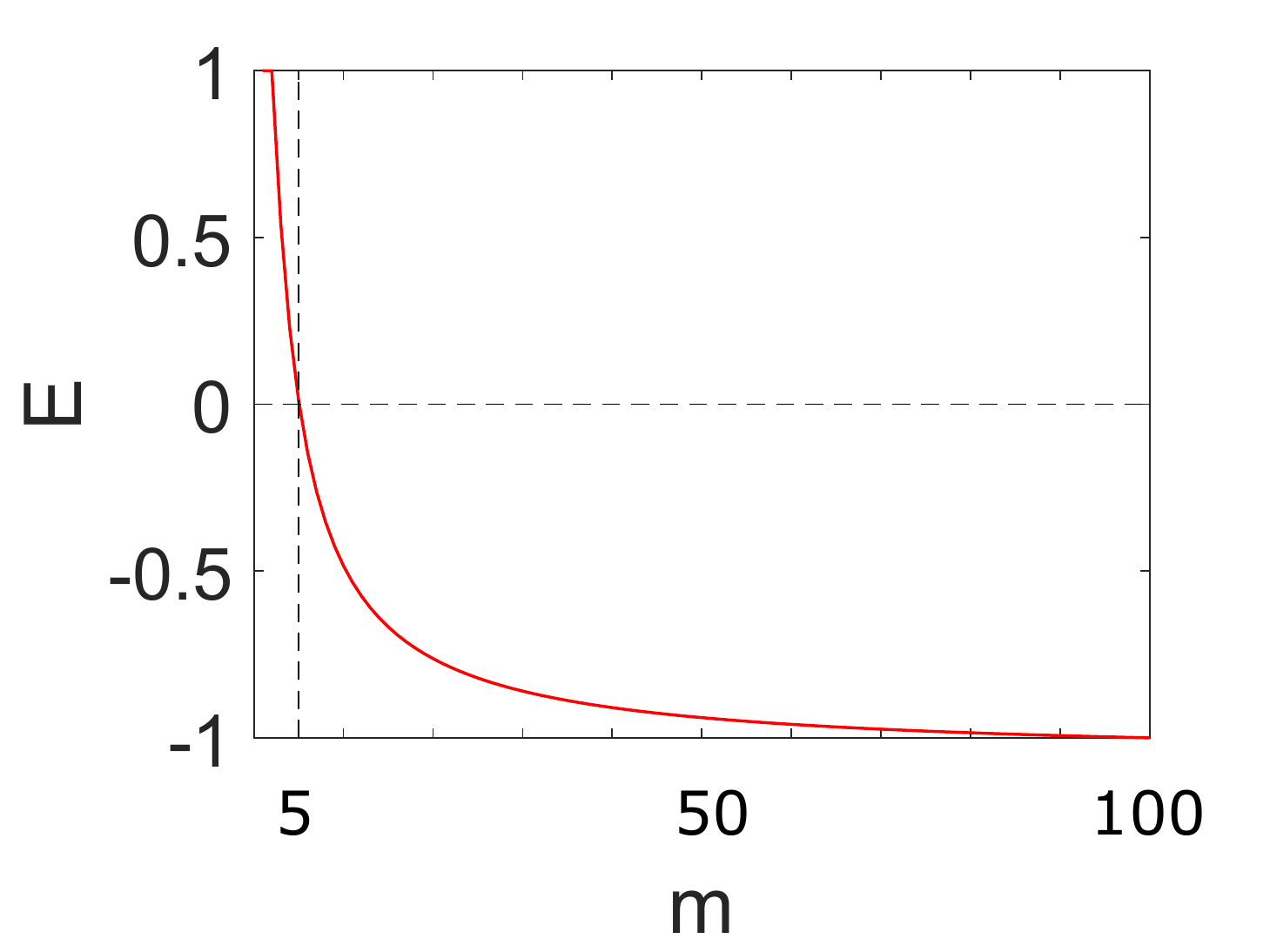}
		\caption{Numerical solution of energy (E) as a function of number of clusters (m), for a network of N=100 nodes.}
		\label{numeric}
	\end{center}
\end{figure}

Thus, summing over all triads yields the upper bound on the energy function:

\begin{eqnarray}
E \geqslant -\sum_{i} \frac{m_{i} \binom {k}{3} - \frac {m_{i}(m_{i}-1)}{2} k(N-m_{i})+ \binom {m_{i}}{3} }{\binom N3}.
\end{eqnarray}
Our results show that the number of stubborn communities formed in a network with similar size is limited (Figure \ref{fig2}). Consider a fixed number of stubborn cliques m, in a network of size N. If all N nodes in the network are uniformly distributed among the $m \sim N/k$ cliques, we have:
\begin{eqnarray}
E \geqslant - \frac{\frac{N}{k} \binom {k}{3} - \frac{N}{2 k}(\frac{N}{k}-1) k(N-\frac{N}{k})+ \binom {\frac{N}{k}}{3} }{\binom N3}.
\end{eqnarray}
In the extreme case $k \rightarrow \infty$, the condition implies that the number of stubborn-communities will be $m \sim 5$. This is consistent with the obtained numerical results, which point at the existence of bounds for the stubborn-community size. Figure. ~\ref{numeric} shows the numerical solution of energy as a function of the number of clusters. As presented in this figure, the number of these stubborn-communities is limited to five communities.
One possible way to find the sign configuration that achieves this upper bound on E is through computational searches: From an initial state in which all the edges are negative, one edge is randomly selected and the sign switches following SBT. This process continues until the network reaches a local minimum of E. For a small network in a zero-energy local minimum, configurations on nodes with several stubborn links are complex. Finding such states on larger networks is more computationally expensive. We test whether such communities are significantly likely to appear in the networks:
\begin{eqnarray}
P_{m,k} \propto \frac{1}{\binom N {k} \frac{k(k-1)}{2} k(N-k)}
\end{eqnarray}
where $P_{m,k}$ is the probability that $k$ nonempty $m$-vertices stubborn-communities form in $G$. Which $P_{m,k}$ is calculated as the product of the probability that no vertex inside the subgraph is connected to any outside vertex ($m(N-m)$) and the probability that the vertices are fully connected, for the same size communities $m = \frac{N}{k}$. With large numbers of nodes, ($N \rightarrow \infty$), $P_{m,k} \rightarrow 0$. Which is consistent with findings of Antal et al. \cite{Antal2005}, who showed that probability to reach jammed states using such local search methods decreases to zero as a function of the network size. Such a probability increases in the presence of stubborn links. According to the computational difficulty in finding such local minima via algorithmic search, we now show how the sign configuration in positive-energy can be seen through a direct construction which is motivated by computational searching for small examples.

Consider a fixed number of stubborn cliques k in which the nodes are equidistributed among the stubborn communities. As the fewest number of edges are in the cliques, hence the least edges are available for inclusion in unbalanced triangles. The simulations show that the stubborn-communities with an equal number of nodes, move the network towards positive energy with a lower stubborn link ratio in comparison with randomly distributed nodes(Figure ~\ref{fig2}B). We now show that this clustered signed complete graph has positive energy. Consider a network with $N$ nodes and $m$ non-overlapping mutually antagonistic cliques of equal sizes $k=N/m$. In the extreme state, all negative links are stubborn and flipping positive links leads to increase unbalanced triads, so energy is increased by $\delta E$:
\begin{eqnarray}
\delta E = \binom m2 (N-k) + \binom {m}{2} [N-2 (k) -2 \binom {k}{2}  k].
\end{eqnarray}
According to the dynamic models for structural balance, the overall number of imbalanced triads cannot increase in an update event, which implies an upper bound on the energy ($E$). Hence, the network evolves into sub-communities with uniformly distributed negative links having the maximum energy and all adjacent sign configurations having lower energies.  This lower bound is related to the constraints $\sum_{m} c_{m} = n $, where $c_{m}$ is the number of nodes in the $m$th stubborn-community. If all the nodes are distributed in $5$ roughly equal-sized groups, then the proportion of unbalanced triangles (and consequently energy) would be maximized. This configuration is the most stressful configuration with the minimum number of stubborn links.
\begin{figure}[h!]
	\begin{center}
		\includegraphics[width=0.5\linewidth]{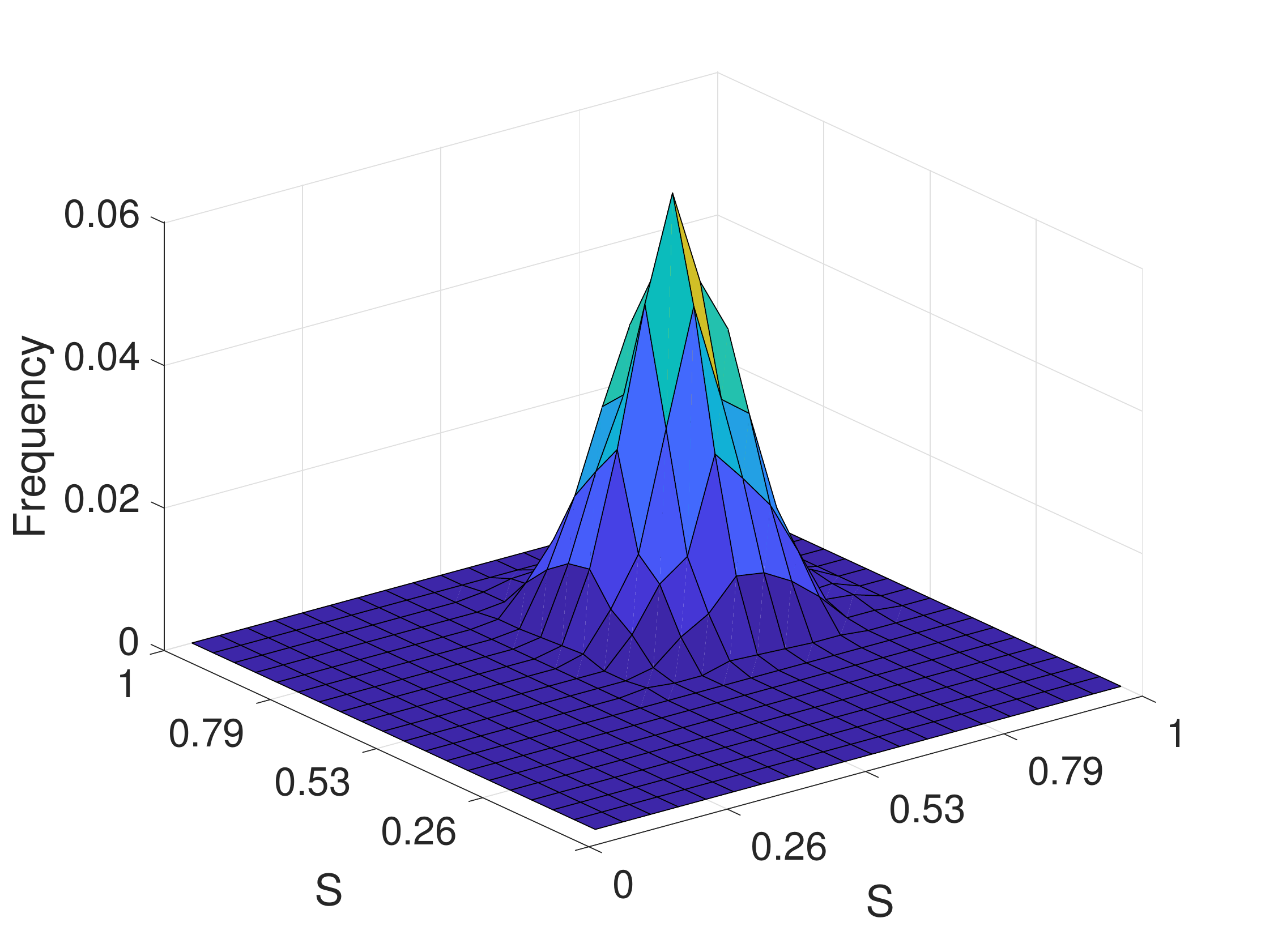}	
	\end{center}
	\caption{S-S distribution of stubborn degree for a network with $N=20$ nodes. The symmetric graphics represent the cases of the minimum number of stubborn links with positive energy.}\label{fig4}
\end{figure}
The correlation between nodes degree based on their number of stubborn links also shows interesting behavior (S in Figure ~\ref{fig4}). Nodes with the same stubbornness degree (defined as the number of stubborn links normalized by the node's degree) are strongly connected to each other. In other words, stubborn nodes tend to connect to those who are similar to them, a condition essential in peer to peer multi-agent competition and cooperation dynamics \cite{Caram2}. The graphics represent the case of the minimum number of stubborn links with positive energy. This correlation supports the idea that emerging stubborn-communities formed by stubborn links lead to positive energies. We argue that if networks with stubborn links are divided into stubborn-communities, the division alone can produce both degree-degree correlations and clustering. This evidence is compatible with the results regarding the stubborn-community forming.
\section{Conclusions}

While structural balance theory shows the tendency of locally interacting agents to avoid conflictual situations, this dynamic changes upon the introduction of stubborn links into social systems. We investigated how instability can emerge as a result of such local stubborn interactions. By means of the structural balance theory, we study the fraction of stubborn links and where to place them in order to maximize the energy. Our results show that these links which are typically few in numbers can alter the equilibrium of the balance dynamics in a fully connected signed network. We observed the tendency to form stubborn communities, such that there is a kind of stubbornness assortativity emerging from the balance dynamics. Specifically, the final state energy of networks does not directly depend on the fraction of stubborn links, it depends on their position in the network, i.e. on the way they topologically assemble into the system, e.g. stubborn communities. To understand how much the fraction of stubborn links could affect the network energy, we numerically found upper and lower bounds to the energy function depending on the number of stubborn links present in the network. We computed the maximum positive energy reachable by a minimum fraction of such links and the minimum negative energy reachable by a maximum fraction of stubborn links. Our analytical results show that in the infinite network size limit, the energy upper bound is reachable with only 5 stubborn communities. This results provide new insights on SBT dynamics in presence of stubborn relations among people, parties or countries.

\bibliographystyle{unsrt}  


\end{document}